\documentstyle[art12,epsf]{article}
\begin{document}
\baselineskip 0.332in
\begin{flushright}
\vskip -2cm
UAHEP971\\ \today \\hep-ph/9702301
\end{flushright}
\begin{center}
\vskip 1cm
{ \Large \bf Light Gluino And Tevatron\\ Dijet Angular Distributions } \\[7mm]
I. Terekhov\footnote{e-mail: terekhov@lepton.ph.ua.edu} \\
{\it Department of Physics and Astronomy\\
The University of Alabama, Tuscaloosa Al 35487.}
\end{center}
\begin{abstract}
We consider the effects of a light gluino on the dijet angular distributions
in $p\overline{p}$ collisions at $\sqrt{s}=1800$ GeV. 
We show that  the slower running of $\alpha_s$
and the production $q\overline{q},gg \rightarrow\tilde{g}\tilde{g}$ do
not significantly alter the normalized differential cross-sections. 
On the other hand, single $\tilde{u},\tilde{d}$ squark production 
$qg\rightarrow \tilde{q}\tilde{g}$ with
subsequent $\tilde{q} \rightarrow q\tilde{g}$ results in dijets whose angular
distributions are dramatically different from that in SM. The CDF data 
on dijet
angular distributions with integrated luminosity of $106 {\rm pb}^{-1}$ exclude
$\tilde{u}$ and $\tilde{d}$ squarks in the mass ranges $ 150 \leq m \leq 650$
and $ 170 \leq m \leq 620$, respectively. We consider lower
energies as well and show that  in a future experimental analysis for
dijet mass below 240 GeV squarks could be observed or ruled out.
\end{abstract}
\thispagestyle{empty}
\setcounter{page}{0}
\newpage

In  recent years, many theoretical proposals beyond the Standard Model have 
been developed; however, Supersymmetry (SUSY) remains one of the best motivated
and attractive models. Extensive experimental searches for SUSY have imposed 
numerous constraints on allowed mass ranges for the super-particles, most of 
the effort being concentrated within the framework of the  Minimal 
Supersymmetric Standard Model (MSSM). 
However, theoretical arguments have been made in 
favor of certain special variants, most notably, the light
gluino scenario where gluino and photino are much lighter than squarks and,
in fact, are (nearly) massless ($\leq 5$GeV). 
The idea acquired a particular interest with the emergence of the 
discrepancy between the measurements of the strong coupling constant at
low energy and at the $M_Z$ scale. The contribution of a neutral colored 
fermion, such as a light gluino, into the $\beta$-function would slow the 
$\alpha_s$ running so as to improve the agreement appreciably \cite{alphas}.
Although recent studies have reported a diminished discrepancy
\cite{alphasaway} and thus necessitate the light gluino to a lesser extent, 
the possibility remains appealing.


The experimental low mass 
window for gluino was first pointed out by the UA1 collaboration \cite{ua1}.
A number of other experiments attempted to search for a light gluino as well;
these include beam dump experiments\cite{bdump}, searches for new neutral 
particles\cite{newpart}, and radiative Upsilon decay
\cite{ups}. Many claimed to have narrowed down the allowed window dramatically.
 Farrar, however, has provided a critical review of these experiments 
\cite{mainfar} and indicated that  most analyses mentioned therein had to 
make assumptions on gluino
decay modes or rely on non-perturbative effects so their claims had to be
weakened. The allowed region in the gluino mass and lifetime
space was given. 
Occasionally, the results of \cite{mainfar} are considered liberal 
in the literature, but they are particularly interesting due to the
possibility of a gluino that is ultra-light ($\leq 1$ GeV). Such a particle 
would
have to be {\it long-lived}, that is, would not decay before 
interacting hadronically in  a detector or beam dump. Possible missing
energy from a gluino weak decay would therefore 
be negligible so the canonical SUSY searches
would have been incapable of detecting it.

Clearly, a low mass long-lived gluino implies a dramatically different
phenomenology. It is expected to form  hadrons such as $R^0$
\cite{mainfar,farr0}
which in turn result in jets in the detector. Consequently, it may be the
jet physics where SUSY with the light gluino is discovered.
Remarkably, such phenomenology has distinct advantages. The principal 
part of the calculations is perturbative, the
resulting cross-sections depend insignificantly on the gluino mass, and
many predictions do not depend on the details of gluino fragmentation.
In the present work, we will continue to study light gluino implications 
for collider jet physics.


Last year Fermilab announced the observation of an excess of high 
transverse energy jet events in $p\overline{p}$ collisions\cite{fnalET}.
Here the light gluino, mostly due to the effect on the $\alpha_s$ running,  
was shown to be able to alleviate the discrepancy noticeably\cite{ct,otherET}, 
although not to fully account for it. At the same time, several possible
suggestions were made within the framework of standard QCD, mostly 
regarding parton density functions (pdf's)\cite{smEt}; 
finally, the CTEQ collaboration has
presented a set of updated pdf's \cite{cteq4} with inclusive jet $E_T$ 
experiments incorporated into the analysis, and the overall consistency 
of the Standard Model has thereby been restored. The new fits required
a significant modification of  the gluon pdf and ignoring a fair
amount of low $E_T$ data, which may certainly be justified given both
theoretical and experimental difficulties in the low energy region 
\cite{cteq4}. However, the agreement
with the CDF data, {\it including} the low $E_T$ region\footnote
{The D0 collaboration has not presented its low energy data.}, is better 
with the light gluino than without it\cite{ct}. 
Since in addition to ordinary partons
a light gluino sea must be present in hadrons \cite{RSpdf,RVpdf}, it would 
be interesting to incorporate the Tevatron $E_T$ data into developing a set 
of pdf's with the light gluino.

Other jet experiments are more sensitive to the low mass gluino.
In this scenario, a single squark can be readily produced
via $qg\rightarrow \tilde{g}\tilde{q}$; see \cite{tc} for the cross-section
as a function of squark mass. 
The subsequent decay 
$\tilde{q}\rightarrow q\tilde{g}$ results in a pair of jets  an
invariant mass close to $m_{\tilde{q}}$. The jet experiments deal
with  dijets, which are defined as the pair of jets with the highest 
$p_T$  in the event subject also to other
selection criteria\cite{CDFdijet,main}. 
Based on the early CDF dijet mass distribution data sample 
of $19 {\rm pb}^{-1}$\cite{CDFdijet}, it was possible to exclude 
$\tilde{u}, \tilde{d}$ squarks between 330 and 440 GeV at 95\% CL\cite{tc}.
In a recent similar analysis of \cite{mudilos}, the excluded region 
has been extended down to 220 GeV and up to 475 GeV.
In the present paper, we address the impact of single squark production on
the distribution of dijet cross-section in the center of mass 
scattering angle $\theta^*$. It should be noted that the work \cite{mudilos}
studies the dijet angular distributions as well. However, the previous analysis
relies on the gluino pdf and considers the $2\rightarrow 2$ process
$q\tilde{g}\rightarrow\tilde{q}\rightarrow q\tilde{g}$. Since the gluino
sea is due to the $g\tilde{g}\tilde{g}$ vertex, all gluino
initiated processes can be conceived of as part of the respective gluon
initiated processes, where the gluino momentum is restricted to be parallel
to the proton beam. Hereinafter we present a complete treatment of the
$2\rightarrow 3$ process
$qg\rightarrow \tilde{q}\tilde{g} \rightarrow \tilde{g}\tilde{g}q$,
where $q=u,d$ \cite{tc}
\footnote{Using the gluino pdf's from
\cite{RSpdf} we have checked that the dijet cross-sections would be similar to
those in \cite{mudilos}. We thank Dr. J. Stirling for furnishing us the
code for pdf evaluation.}.

In the standard QCD, the jet (and dijet) angular distribution is sharply
peaked in the forward direction due to the t-channel exchange of a (nearly) 
massless parton. In SUSY with light gluino, several new effects are added. The
modification of $\alpha_s$ running is expected to be relatively unimportant
as the pattern of dijet angular distribution is roughly the same for
the whole energy range involved. Two more processes of
a gluino pair production must also be considered:
$q\overline{q}, gg \rightarrow \tilde{g}\tilde{g}$; but again, 
the situation is not expected to change
qualitatively as the resulting picture will be largely determined by
the still massless t-channel gluinos. 
On the other hand, the production of almost any 
heavy particle which decays somewhat isotropically in its rest frame, 
such as a squark, 
flattens the observed jet angular distribution and can be sensed
even with a small production cross-section.

Recently, CDF presented
the $106 {\rm pb}^{-1}$ data sample on the dijet angular distribution measurements
\cite{main}. Instead of $\cos \theta^*$, a more convenient variable 
$\chi \stackrel{\rm def}{=} (1 + \left | \cos \theta^* \right | )
/ (1 - \left | \cos \theta^* \right | )$ was used for differential 
cross-sections as $d\sigma / d \chi$ is relatively flat for background
processes. Here  $\cos \theta^*$ is actually defined through the observable 
jet pseudorapidities as $\tanh\frac{\eta_a-\eta_b}{2}$ and has a meaning
of cosine CMS scattering angle only for a pair of jets with balanced
$p_T$. New processes with flatter $d\sigma / d \cos \theta^*$ would result in
an excess of low $\chi$ ($\approx 1$) events. It is important to note that
the normalized  cross-section
$ (1/\sigma_{tot}) d\sigma / d\chi$ 
has little sensitivity to higher order corrections
and the scale at which pdf's and $\alpha_s$ are evaluated \cite{main}.

\begin{figure}
\hskip 2cm
\epsfxsize=5in \epsfysize=5in 
\epsfbox{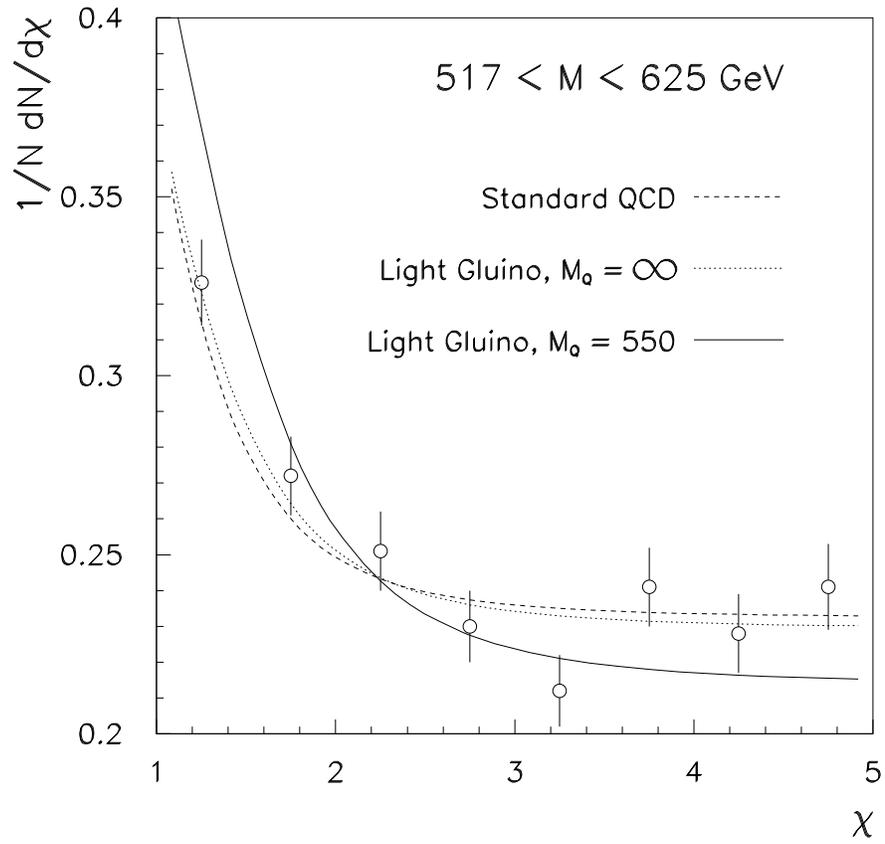}
\caption{Production of a $\tilde{u}$ squark ($m_{\tilde{u}}=550$ Gev) 
and angular 
distributions of dijets with the mass $517 \leq M \leq 625$ GeV.
The LO predictions of Standard Model and SUSY with light 
gluinos and infinitely heavy squarks are also shown.}
\label{p517625}
\end{figure}

We evaluate the squared amplitude for
$qg\rightarrow \rightarrow \tilde{g}\tilde{g}q$ \cite{tc}
and  perform a multidimensional Monte-Carlo integration over the phase space.
In computing the $\chi$ for an event we select the two
highest $p_T$ jets and use the CDF pseudorapidity cuts 
$\left | \eta_{a,b} \right | < 2.17$, as well as constrain the dijet
invariant mass to the same mass bins as in \cite{main}. We take
the squark width to be 
$ \Gamma(\tilde{q}\rightarrow q\tilde{g}) =2/3\alpha_s m_{\tilde{q}}
\approx \Gamma_{tot}$ thus neglecting non-hadronic modes. 
We also sum over left/right squark production neglecting the possible mass 
splitting between these states.
Since there is no interference between diagrams containing $\tilde{q}_L$ and 
$\tilde{q}_R$, in the case of $M_R-M_L$ mass splitting large relative to the
$\tilde{q}$ width one simply divides the cross-sections by 2 to
get separate cross-sections.
We have taken the gluino mass to be 100 MeV. We employ
the CTEQ3L (leading order QCD fit without the Tevatron jet $E_T$ data) 
pdf's\cite{cteq3}
and evaluate them at  half of the transverse energy of the leading jet.
The strong coupling constant is evaluated
at the same scale using SUSY RGE equations with $\alpha_s(M_Z)=0.12$.
We also evaluate in the lowest order 
all relevant background $2\rightarrow 2$ processes, including
gluino pair production. The latter amount
for only 6-10\% of the standard processes\cite{ct,otherET,mudilos}. 
The resulting normalized
distributions along with the actual data (statistical errors only) 
are illustrated in Fig. \ref{p517625} for the dijet mass range of 
$517 \leq M \leq 625$ GeV and a $m_{\tilde{u}}\approx$ 550 GeV squark.
As expected, the distribution for SUSY with light gluinos but  heavy 
decoupled squarks is rather close to that for  standard QCD, 
whereas the addition of squark production is distinctly inconsistent 
with the data. In particular, the excess of dijet events in this mass bin 
\cite{CDFdijet} cannot be associated with squarks in the light gluino
scenario. 

For a more detailed analysis, we use the CDF data on $R_{\chi}$, which is
defined for  each mass bin as the ratio of the number of events with
$\chi \leq 2.5$ to that with $\chi > 2.5$. The proprietary choice of 2.5
for the
``pivot point'' was related to the CDF search for
the quark compositeness, but it is suitable for our procedure as well.
Unlike $d\sigma/d\chi$, the $R_{\chi}$ data is provided with systematic
errors and their correlation (full correlation is suggested in \cite{main}
among systematic errors for the five mass bins). We add the systematic
error matrix in quadrature to the statistical error matrix 
(zero correlation) in computing the full covariance matrix $V_{ij}$. 
We then evaluate
\[
\chi^2=\sum_{i,j=1}^{i,j=5}\Delta R_i\Delta R_j (V^{-1})_{ij}
\]
where $\Delta R_i(m_{\tilde{q}}) = R^{SUSY}_i(m_{\tilde{q}}) - R^{data}_i$, 
for different squark
masses. Here $R^{SUSY}(m_{\tilde{q}})$ is the value of $R_{\chi}$
in SUSY with light gluinos including squark effects.
Assuming different masses of $\tilde{u}, \tilde{d}$, we perform
the $\chi^2$ analysis separately for the two squark flavors corresponding to
valence quarks in the proton. We have found that $\tilde{u}$ squarks in the
mass range 150 -- 650 GeV and  $\tilde{d}$ squarks in the range 170 -- 620 GeV
are incompatible in the $\chi^2$ sense with the CDF data at the 95\% CL,
in rough agreement with \cite{mudilos}.

Although the CDF analysis of their dijet angular distribution data has
been primarily oriented  toward possible new physics at  high energies 
($\sim 1$ Tev), such as  quark substructure,
lower energies may as well
be interesting  for analyses such as ours. In the Supergravity SUSY models 
\cite{sugra}, the light gluino implies low masses of squarks, below a few 
hundred GeV, if a light stop contributes appreciably into the $R_b$
ratio \cite{Rb}. More importantly, the CDF data on jet transverse energy 
distributions may suggest $\tilde{u}, \tilde{d}$ quarks near $\approx 106$ 
GeV \cite{tc}. We hope that if the gluino is light, Supersymmetry will
be observed at these energies.
It is therefore imperative to make predictions about
the effects on dijet angular data of squarks in the 50 -- 200 GeV range
\footnote{Note that the Tevatron searches \cite{heavyFNAL} assume
 the missing energy signature of squarks and therefore their results
are not applicable to the light gluino scenario.}.
Special attention to this low mass region is another major difference
between the present and earlier analyses.

\begin{figure}
\hskip 2cm
\epsfxsize=5in \epsfysize=5in \epsfbox{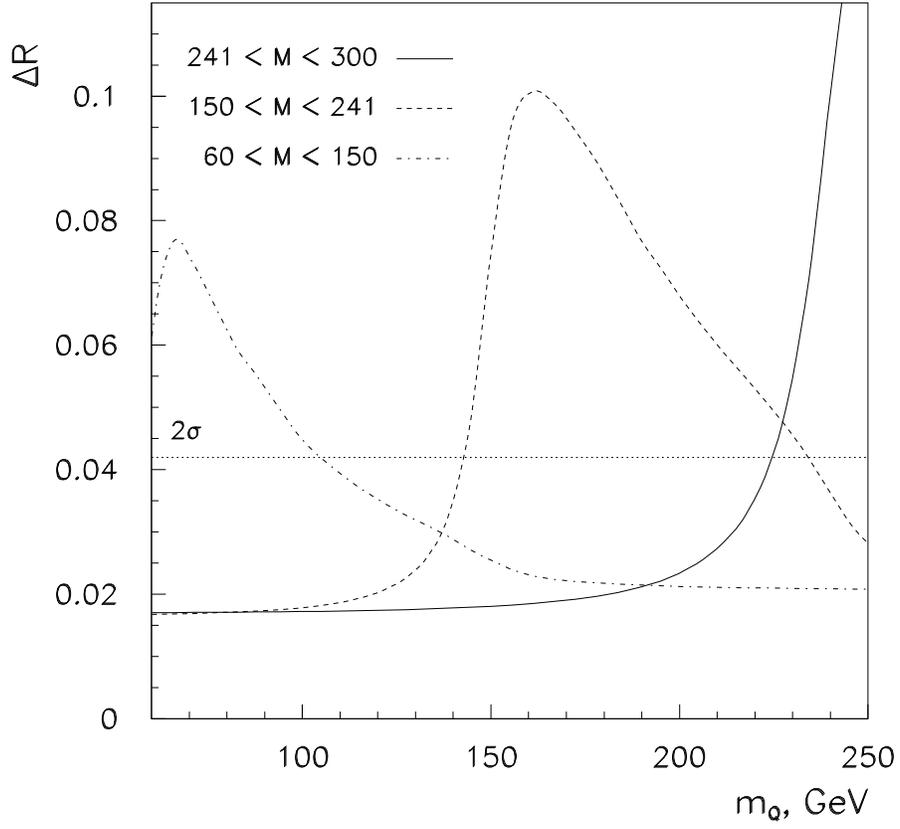}
\caption{Predictions for $\Delta R^{\prime}=R^{SUSY}-R^{QCD}$ for three
dijet mass ranges. The dotted line is the projected $2\sigma$ level, where 
$\sigma=0.021$ is an example taken from the experimental uncertainty in 
$R_{\chi}$ in the lowest
published dijet mass bin, $241\leq M \leq 300$ Gev.}
\label{predict}
\end{figure}

In addition to the five published CDF dijet mass bins, we consider two more 
bins, $60 \leq M \leq 150$ and 
 $150 \leq M \leq 241$ GeV and compute the expected deviation
from the standard QCD
$\Delta R^{\prime}_i (m_{\tilde{q}})=R^{SUSY}_i(m_{\tilde{q}})-R_i^{QCD}$,
where, as above, $i$ distinguishes dijet mass bins.
The result is plotted in Fig.\ref{predict} for the two suggested mass bins 
along with that for the lowest 
actual CDF mass bin of  $241 \leq M \leq 300$ GeV. We compare 
$\Delta R^{\prime}$ to the  uncertainty in the measured $R_{\chi}$ in
the $241 \leq M \leq 300$ bin, $\sigma=0.021$, by plotting the 
$2\sigma$ level line.
Clearly, if the CDF is able to determine $R_{\chi}$ for  dijet masses such
as in Fig. \ref{predict} with an error of $\stackrel{<}{\sim}0.02$, 
squarks below
$\approx\! 150$ GeV, will be observed or excluded at the 95\% CL.  
In the latter case, at least some Supergravity related models with light 
gluino will be ruled out even without upgrading the Tevatron energy.

To recapitulate, we have studied the impact of the light gluino on dijet
angular distributions at the Tevatron. We have shown that the major effect
is associated with the single squark production. 
The current  CDF data with 106 ${\rm pb}^{-1}$ integrated luminosity are
inconsistent with $170 \leq m_{\tilde{q}} \leq 650$ Gev
squarks in the light gluino case.
The same or better precision data for dijet masses below 240 GeV will
allow one to discover the squarks near $\approx$ 100 GeV, which may be
suggested by the $R_b$ anomaly or by the CDF inclusive jet $E_T$ distributions.
Since any SUSY theory can only be attractive if it exists at  the electroweak 
scale, the dijet angular measurements, depending on luminosity,
 may prove crucial tests  for the light gluino hypothesis.

I am thankful to Dr. Clavelli at the University of Alabama for stimulating
discussions  without which this work would not have been possible.

\end{document}